\documentclass{LMCS}
\usepackage{here,amsfonts,pstricks,pst-node,
pst-plot,pst-coil,hyperref}

\def\sqr#1#2{{\vcenter{\hrule height .#2pt
         \hbox{\vrule width.#2pt height#1pt \kern#1pt
             \vrule width.#2pt}
         \hrule height.#2pt}}}

\def\bigsquare{\mathchoice\sqr76\sqr76\sqr{2.1}3\sqr{1.5}3}

\def\doi{2 (5:3) 2006}
\lmcsheading%
{\doi}
{1--7}  
{}
{}
{Jun.~\phantom{0}9, 2006}
{Nov.~07, 2006}
{}   

\begin{document}
\title{The Completeness of Propositional Resolution\\
A Simple and Constructive Proof}
\author[J.~Gallier]{Jean Gallier}
\address{
CIS Department \\
University of Pennsylvania\\
Philadelphia, PA 19104, USA}
\email{jean@cis.upenn.edu}
\begin{abstract}
\noindent
It is well known that the resolution method (for propositional 
logic) is complete. However,  completeness proofs found
in the literature use  an argument by contradiction
showing that if a set of clauses is unsatisfiable, then it must
have a resolution refutation. As a consequence, none of these proofs
actually gives an algorithm  for producing a resolution refutation
from an unsatisfiable set of clauses.
In this note, we give a simple  and constructive
proof of the completeness of propositional resolution
which consists of an algorithm together with 
a proof of its correctness.
\end{abstract}

\maketitle

\section{Introduction}
\label{sec1}

The resolution method for (propositional) logic
due to J.A. Robinson \cite{Robinson65} (1965)
is well-known to be a sound and complete procedure
for checking the unsatisfiability of a set of clauses.
However, it appears that the completeness proofs that can be found in the 
literature (for instance,
Chang and Lee \cite{ChangLee}, Lewis and Papadimitriou
\cite{LewisPapa}, Robinson \cite{Robinson79})
are existence proofs that proceed by contradiction
to show  that if a set of clauses is unsatisfiable, then it must have
a resolution refutation because otherwise 
a satisfying assignment can be obtained.
In particular, none of these proofs
yields (directly) an algorithm  producing a resolution refutation
from an unsatisfiable set of clauses.
In that sense, these proofs  are nonconstructive.
In Gallier \cite{Gallbooklogic} (1986),
we gave a completeness proof based on an algorithm for converting 
a Gentzen-like proof (using sequents) into a resolution DAG (see Chapter 4).
Such a method is more constructive than the others but,
we found later on that it is possible to give a simple
and constructive proof  of the completeness of
propositional resolution which consists of an algorithm
together with a proof of its correctness. 
This algorithm and its correctness
are the object of this note.

It should be noted that  Judith Underwood
gave other constructive proof procedures
in her Ph.D. thesis,  notably 
for the intuitionistic propositional calculus
\cite{Underwood}.

\section{Review of Propositional Resolution}
\label{sec2}

Recall that a {\it literal\/}, $L$, is either a
propositional letter, $P$, or the negation, $\neg P$, of
a propositional letter. A {\it clause\/} is a 
finite set of literals, $\{L_1, \ldots, L_k\}$,
interpreted as the disjunction $L_1\lor \cdots \lor L_k$
(when $k = 0$, this is the empty clause denoted $\bigsquare$).
A set of clauses, $\Gamma = \{C_1, \ldots, C_n\}$, is interpreted as
the conjunction $C_1\land\cdots\land C_n$. For short, we
write $\Gamma = C_1, \ldots, C_n$.

The {\it resolution method\/} (J.A. Robinson \cite{Robinson65}) 
is a procedure for
checking whether a set of clauses, $\Gamma$,  is unsatisfiable.
The resolution method consists in building a certain kind of labeled DAG
whose leaves are labeled with clauses in $\Gamma$ and whose
interior nodes are labeled according to the {\it resolution rule\/}.
Given two clauses $C =  A\cup\{P\}$ and $C' = B\cup\{\neg P\}$ (where
$P$ is a propositional letter, $P\notin A$ and $\neg P \notin B$), 
the {\it resolvent of $C$ and $C'$\/}
is the clause
\[
R = A\cup B
\]
obtained by cancelling out $P$ and $\neg P$. 
A {\it resolution DAG for $\Gamma$\/} is a DAG whose
leaves are labeled with  clauses from $\Gamma$ and such that every interior
node $n$ has exactly two predecessors, $n_1$ and $n_2$ so that 
$n$ is labeled with the resolvent of the clauses labeling $n_1$ and $n_2$.
In a resolution  step involving the nodes, $n_1, n_2$ and $n$, as above,
we say that the two clauses $C$ and $C'$ labeling the nodes $n_1$ and
$n_2$ are the {\it parent clauses\/} of the resolvent clause, $R$, labeling
the node $n$. In a resolution DAG, $D$, a clause, $C'$ is said
to be a {\it descendant\/} of a clause, $C$, iff
there is a (directed) path from some node labeled with $C$ to a node
labeled with $C'$.
A {\it resolution refutation for $\Gamma$\/} is a resolution
DAG with a single root whose label is the empty clause.
(For more details on the resolution method, resolution DAGs, etc., 
one may consult Gallier \cite{Gallbooklogic}, Chapter 4, or any of the
books cited in Section \ref{sec1}.)

Here is an example of a resolution refutation for the set of clauses
\[
\Gamma = \{\{P, Q\}, \{P, \neg Q\}, \{\neg P, Q\}, \{\neg P, \neg Q\}\}
\]
shown in Figure \ref{fig0}:
\begin{center}
\begin{figure}[H]
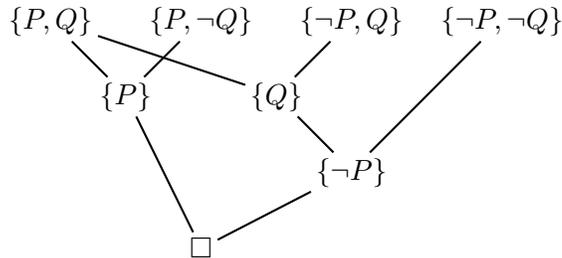

\pspicture(0, 0)(7, 3)
\rput(0, 3){\Rnode{a}{$\{P, Q\}$}}
\rput(2, 3){\Rnode{b}{$\{P, \neg Q\}$}}
\rput(4, 3){\Rnode{c}{$\{\neg P, Q\}$}}
\rput(6, 3){\Rnode{d}{$\{\neg P, \neg Q\}$}}
\rput(1, 2){\Rnode{e}{$\{P\}$}}
\rput(3, 2){\Rnode{f}{$\{Q\}$}}
\rput(4, 1){\Rnode{g}{$\{\neg P\}$}}
\rput(2, 0){\Rnode{h}{$\square$}}
\ncline[nodesep=2pt]{a}{e}
\ncline[nodesep=2pt]{b}{e}
\ncline[nodesep=2pt]{a}{f}
\ncline[nodesep=2pt]{c}{f}
\ncline[nodesep=2pt]{e}{h}
\ncline[nodesep=2pt]{d}{g}
\ncline[nodesep=2pt]{f}{g}
\ncline[nodesep=2pt]{g}{h}
\endpspicture
\caption{A Resolution Refutation}
\label{fig0}
\end{figure}
\end{center}

\section{Completeness of Propositional Resolution: \\
An Algorithm and
its Correctness}
\label{sec3}
Let $\Gamma$ be a set of clauses. Thus, $\Gamma$ is either
the empty clause, $\bigsquare$, or it is a conjunction
of clauses, $\Gamma = C_1, \ldots, C_n$. 
We define the
{\it complexity\/}, $c(C)$, of a clause, $C$, as the number
of disjunction symbols in $C$; i.e., if $C$ consists of a single literal
(i.e., $C = \{L\}$, for some literal, $L$), then
$c(C) = 0$, else if
$C = \{L_1, \ldots,  L_m\}$ (with $m \geq 2$) where the $L_i$'s are literals, 
then $c(C) = m - 1$ (we also set $c(\bigsquare) = 0$).
If $\Gamma$ is a conjunction of clauses,
$\Gamma = C_1, \ldots, C_n$, then we set
\[
c(\Gamma) = c(C_1) + \cdots + c(C_n ).
\]

We now give a recursive algorithm, {\tt buildresol},
for constructing a resolution DAG
from any set of clauses and then
prove its correctness, namely, that if the input
set of clauses is unsatisfiable, then the output
resolution DAG is a resolution refutation.
This  establishes the
completeness of propositional resolution constructively. 

Our algorithm makes use of two functions, {\tt percolate},
and {\tt graft}.

\noindent
{\bf 1. The function {\tt percolate}$(D, A, L)$}

The inputs are: a resolution DAG, $D$, some
selected leaf of $D$ labeled with a clause, $A$, and some literal, $L$.
This function adds the literal $L$ to the clause $A$ to form the 
clause $A\cup \{L\}$ and then ``percolates'' $L$  down to the root
of $D$. 
More precisely, we construct
the resolution DAG, $D'$, whose underlying unlabeled DAG
is identical to $D$, as follows:
Since $D$ and $D'$ have the same unlabeled DAG 
we  refer to two nodes of $D$ of $D'$ as
{\it corresponding nodes\/} if they are identical
in the underlying unlabeled DAG.
Consider any resolution step of $D$.
If both parent clauses are not descendants of the premise $A$,
then the corresponding resolution step of $D'$ is the same.
If the parent clauses in $D$ are $C$ and $C'$
where $C'$ is a descendant of
the premise $A$ (resp.  $C$ is a descendant of the premise $A$)
and if $R$ is the resolvent ot $C$ and $C'$ in $D$, then
the corresponding parent nodes in $D'$ are labeled
with $C $ and  $C' \cup\{L\}$ and their resolvent node
with $R\cup \{L\}$ (resp. the corresponding parent nodes in $D'$ 
are labeled with $C \cup\{L\}$ and $C'$ and their resolvent node 
wih  $R\cup \{L\}$). 
If both  parent clauses $C$ and $C'$ in $D$ are descendant of the premise
$A$, then the corresponding parent nodes in $D'$ are labeled
with $C\cup\{L\}$ and $C' \cup\{L\}$ and their resolvent node
with $R\cup \{L\}$.

Observe that if $\Delta\cup \{A\}$ is the set of premises of
$D$, then $\Gamma = \Delta\cup \{A \cup \{L\}\}$
is the set of premises of {\tt percolate}$(D, A, L)$.

\begin{center}
\begin{figure}[H]
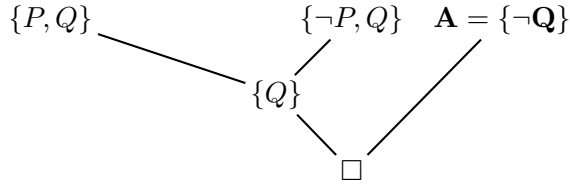

\pspicture(0, 1)(7, 3)
\rput(0, 3){\Rnode{a}{$\{P, Q\}$}}
\rput(4, 3){\Rnode{c}{$\{\neg P, Q\}$}}
\rput(6, 3){\Rnode{d}{$\mathbf{A} = \{\neg \mathbf{Q}\}$}}
\rput(3, 2){\Rnode{f}{$\{Q\}$}}
\rput(4, 1){\Rnode{g}{$\square$}}
\ncline[nodesep=2pt]{a}{f}
\ncline[nodesep=2pt]{c}{f}
\ncline[nodesep=2pt]{d}{g}
\ncline[nodesep=2pt]{f}{g}
\endpspicture
\caption{Resolution DAG $D$}
\label{fig1}
\end{figure}
\end{center}

For example, if $D$ is the resolution DAG shown in Figure \ref{fig1}
(in fact, a resolution refutation)


\noindent
then  adding $L = \neg P$ to $A = \{\neg Q\}$  in $D$ yields
the resolution DAG $D'$ produced by {\tt percolate}$(D, A, L)$
shown in Figure \ref{fig2}:

\begin{center}
\begin{figure}[H]
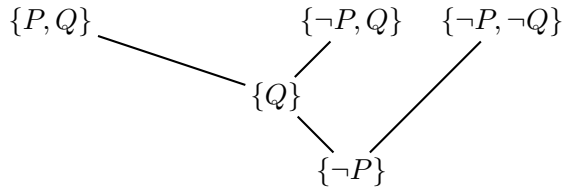

\pspicture(0, 1)(7, 3)
\rput(0, 3){\Rnode{a}{$\{P, Q\}$}}
\rput(4, 3){\Rnode{c}{$\{\neg P, Q\}$}}
\rput(6, 3){\Rnode{d}{$\{\neg P, \neg Q\}$}}
\rput(3, 2){\Rnode{f}{$\{Q\}$}}
\rput(4, 1){\Rnode{g}{$\{\neg P\}$}}
\ncline[nodesep=2pt]{a}{f}
\ncline[nodesep=2pt]{c}{f}
\ncline[nodesep=2pt]{d}{g}
\ncline[nodesep=2pt]{f}{g}
\endpspicture
\caption{Resolution DAG $D' = {\tt percolate}(D, A, L)$}
\label{fig2}
\end{figure}
\end{center}

\noindent
{\bf 2. The function {\tt graft}$(D_1, D_2)$}

Its inputs are two resolution DAGs, $D_1$ and $D_2$,
where the clause, $C$,  labeling the root of
$D_1$ is identical to one of the premises
of $D_2$. Then, this function 
combines $D_1$ and $D_2$ by connecting the links to
the premise labeled $C$ in $D_2$ to the root of $D_1$, 
also labeled $C$, obtaining the resolution DAG
{\tt graft}$(D_1, D_2)$.

\begin{center}
\begin{figure}[H]
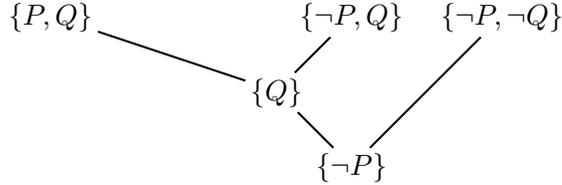

\pspicture(0, 1)(7, 3)
\rput(0, 3){\Rnode{a}{$\{P, Q\}$}}
\rput(4, 3){\Rnode{c}{$\{\neg P, Q\}$}}
\rput(6, 3){\Rnode{d}{$\{\neg P, \neg Q\}$}}
\rput(3, 2){\Rnode{f}{$\{Q\}$}}
\rput(4, 1){\Rnode{g}{$\{\neg P\}$}}
\ncline[nodesep=2pt]{a}{f}
\ncline[nodesep=2pt]{c}{f}
\ncline[nodesep=2pt]{d}{g}
\ncline[nodesep=2pt]{f}{g}
\endpspicture
\caption{Resolution DAG $D_1$}
\label{fig3}
\end{figure}
\end{center}

\begin{center}
\begin{figure}[H]
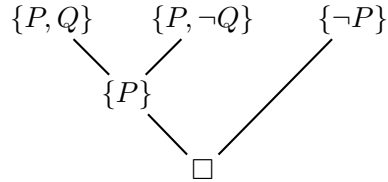

\pspicture(0, 1)(3, 3)
\rput(0, 3){\Rnode{a}{$\{P, Q\}$}}
\rput(2, 3){\Rnode{b}{$\{P, \neg Q\}$}}
\rput(4, 3){\Rnode{c}{$\{\neg P\}$}}
\rput(1, 2){\Rnode{e}{$\{P\}$}}
\rput(2, 1){\Rnode{h}{$\square$}}
\ncline[nodesep=2pt]{a}{e}
\ncline[nodesep=2pt]{b}{e}
\ncline[nodesep=2pt]{e}{h}
\ncline[nodesep=2pt]{c}{h}
\endpspicture
\caption{Resolution DAG $D_2$}
\label{fig4}
\end{figure}
\end{center}

For example, if $D_1$  and $D_2$ are the resolution 
refutation DAGs shown in Figure \ref{fig3} and Figure \ref{fig4}
\vskip 0.5cm

\begin{center}
\begin{figure}[H]
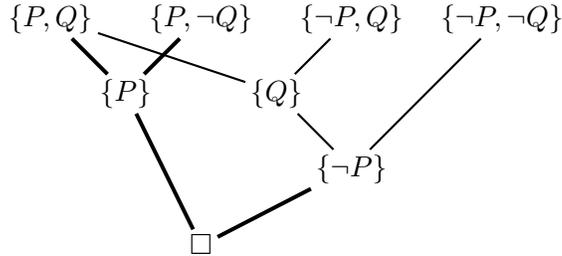

\pspicture(0, 0)(7, 3)
\rput(0, 3){\Rnode{a}{$\{P, Q\}$}}
\rput(2, 3){\Rnode{b}{$\{P, \neg Q\}$}}
\rput(4, 3){\Rnode{c}{$\{\neg P, Q\}$}}
\rput(6, 3){\Rnode{d}{$\{\neg P, \neg Q\}$}}
\rput(1, 2){\Rnode{e}{$\{P\}$}}
\rput(3, 2){\Rnode{f}{$\{Q\}$}}
\rput(4, 1){\Rnode{g}{$\{\neg P\}$}}
\rput(2, 0){\Rnode{h}{$\square$}}
\ncline[nodesep=2pt,linewidth=1.5pt]{a}{e}
\ncline[nodesep=2pt,linewidth=1.5pt]{b}{e}
\ncline[nodesep=2pt]{a}{f}
\ncline[nodesep=2pt]{c}{f}
\ncline[nodesep=2pt,linewidth=1.5pt]{e}{h}
\ncline[nodesep=2pt]{d}{g}
\ncline[nodesep=2pt]{f}{g}
\ncline[nodesep=2pt,linewidth=1.5pt]{g}{h}
\endpspicture
\caption{Resolution DAG {\tt graft}($D_1, D_2)$}
\label{fig5}
\end{figure}
\end{center}

%
we obtain the resolution DAG in Figure \ref{fig5}

where the edges coming from $D_2$ are indicated with thicker lines.
The algorithm {\tt buildresol} is shown below.

\noindent
{\bf 3. The algorithm {\tt buildresol}$(\Gamma)$}

The input to {\tt buildresol} is a set of clauses, $\Gamma$.

\medskip\noindent
{\bf function} {\tt buildresol}$(\Gamma)$

\smallskip\noindent
{\bf begin}

\smallskip\noindent
{\bf if}  all clauses in $\Gamma$ are literals {\bf then}

\smallskip
{\bf if} $\Gamma$ contains complementary literals $L$ and $\neg L$,

\smallskip
{\bf then} return a resolution refutation with leaves $L$ and $\neg L$

\smallskip
{\bf else} {\bf abort}

\smallskip
{\bf endif}

\smallskip\noindent
{\bf else}
select any nonliteral clause, $C$, in $\Gamma$ and select any literal, $L$,
in $C$;

\smallskip\noindent
let $C = A\cup \{L\}$;
let $\Gamma = \Delta \cup \{C\}$; 

\smallskip\noindent
$D_1 = {\tt buildresol}(\Delta\cup \{A\})$;
$D_2 =  {\tt buildresol}(\Delta\cup \{L\})$;
$D_1' = {\tt percolate}(D_1, A, L)$;

\smallskip
{\bf if} $D_1'$ is a resolution DAG 

{\bf then} 
return $D_1'$

\smallskip
{\bf else}
$D = {\tt graft}(D_1', D_2)$;
return $D$

\smallskip
{\bf endif}

\smallskip\noindent
{\bf endif}

\smallskip\noindent
{\bf end} 

Finally, we prove the correctness of our recursive algorithm {\tt buildresol}.

\begin{thm}
\label{compP} 
For every conjunction of clauses, $\Gamma$, if
$\Gamma$ is unsatisfiable, then the algorithm
{\tt builresol} outputs a
resolution refutation for $\Gamma$.
Therefore, propositional resolution is complete.
\end{thm}

\proof
We prove the correctness of the algorithm {\tt buildresol} 
by induction on $c(\Gamma)$.
Let $\Gamma = C_1, \ldots, C_n$. We may assume
$\Gamma\not= \bigsquare$, since the case $\Gamma = \bigsquare$ is trivial.
We proceed by induction on $c(\Gamma)$. 

If $c(\Gamma) = 0$, then every clause, $C_i$, contains a single literal
and if $\Gamma$ is unsatisfiable, then there must be two complementary
clauses, $C_i = \{P\}$ and $C_j = \{\neg P\}$, in $\Gamma$. Thus,
we instantly get a resolution refutation by applying the
resolution rule to $\{P\}$ and $\{\neg P\}$.

Otherwise, $c(\Gamma) > 0$, so there is some
clause in $\Gamma$ that contains at least two literals.
Pick any such clause, $C$, and pick any literal, $L$, in $C$.
Write $C = A\cup \{L\}$ with  $A\not= \bigsquare$ and
write $\Gamma = \Delta,  C$ ($\Delta$ can't be empty
since $\Gamma$ is unsatisfiable).
As $\Gamma = \Delta, A\cup \{L\}$ is unsatisfiable, {\it both\/} 
$\Delta,  A$ and $\Delta,  L$ must be unsatisfiable.
However, observe that
\[
c(\Delta,  A) < c(\Gamma)
\quad\hbox{and}\quad
c(\Delta,  L) < c(\Gamma).
\]
Therefore, by the induction hypothesis, the algorithm
{\tt buildresol} produces
two resolution refutations, $D_1$ and $D_2$, with sets of 
premises $\Delta, A$ and $\Delta, L$, respectively.
Now, consider the resolution DAG, $D_1' = {\tt percolate}(D_1, A, L)$,
obtained from $D_1$ by adding $L$ to the clause $A$ and letting $L$
percolate down to the root. 

Observe that in $D_1'$, every clause that is a descendant of
the premise $A\cup \{L\}$ is of the form $C\cup \{L\}$, where
$C$ is the corresponding clause in $D_1$.
Therefore, the root of the new DAG $D_1'$ obtained from $D_1$ 
is either labeled $\bigsquare$
(this may happen when the other clause in a resolution step 
involving a descendent of the clause $A$
already contains $L$) or $L$.
In the first case, $D_1'$ is already a resolution refutation for
$\Gamma$ and we are done. In the second case,
we can combine $D_1'$ and $D_2$ using {\tt graft}$(D_1', D_2)$
since the root of $D_1'$ is also labeled $L$, one of the premises
of $D_2$. Clearly, we obtain a resolution refutation for $\Gamma$.\qed

As an illustration of our algorithm, consider the set of clauses
\[
\Gamma = \{\{P, Q\}, \{P, \neg Q\}, \{\neg P, Q\}, \{\neg P, \neg Q\}\}
\]
as above and pick $C = \{\neg P, \neg Q\}$, $L = \neg P$ and
$A = \{\neg Q\}$.
After the two calls \\
{\tt buildresol}$(\Delta\cup \{A\})$
and  {\tt buildresol}$(\Delta\cup \{L\})$,  
we get the resolution refutations $D_1$ shown in Figure \ref{fig6}: 

\begin{center}
\begin{figure}[H]
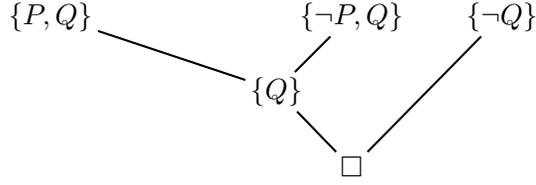

\pspicture(0, 1)(7, 3)
\rput(0, 3){\Rnode{a}{$\{P, Q\}$}}
\rput(4, 3){\Rnode{c}{$\{\neg P, Q\}$}}
\rput(6, 3){\Rnode{d}{$\{\neg Q\}$}}
\rput(3, 2){\Rnode{f}{$\{Q\}$}}
\rput(4, 1){\Rnode{g}{$\square$}}
\ncline[nodesep=2pt]{a}{f}
\ncline[nodesep=2pt]{c}{f}
\ncline[nodesep=2pt]{d}{g}
\ncline[nodesep=2pt]{f}{g}
\endpspicture
\caption{Resolution DAG $D_1= {\tt buildresol}(\Delta\cup \{A\})$}
\label{fig6}
\end{figure}
\end{center}
and $D_2$ shown in Figure \ref{fig7}:

\begin{center}
\begin{figure}[H]
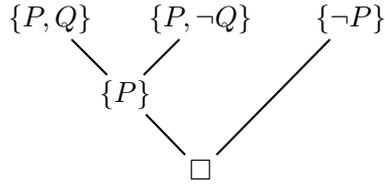

\pspicture(0, 1)(3, 3)
\rput(0, 3){\Rnode{a}{$\{P, Q\}$}}
\rput(2, 3){\Rnode{b}{$\{P, \neg Q\}$}}
\rput(4, 3){\Rnode{c}{$\{\neg P\}$}}
\rput(1, 2){\Rnode{e}{$\{P\}$}}
\rput(2, 1){\Rnode{h}{$\square$}}
\ncline[nodesep=2pt]{a}{e}
\ncline[nodesep=2pt]{b}{e}
\ncline[nodesep=2pt]{e}{h}
\ncline[nodesep=2pt]{c}{h}
\endpspicture
\caption{Resolution DAG $D_2 = {\tt buildresol}(\Delta\cup \{L\})$}
\label{fig7}
\end{figure}
\end{center}

When we add $L = \neg P$ to $A = \{\neg Q\}$  in $D_1$, we
get the resolution DAG \\
$D_1' = {\tt percolate}(D_1, A, L)$ shown in Figure \ref{fig8}:

\begin{center}
\begin{figure}[H]
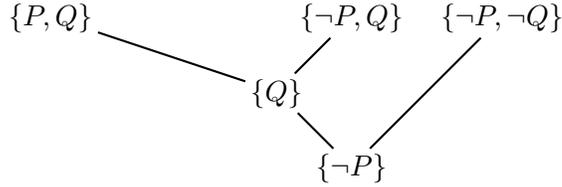

\pspicture(0, 1)(7, 3)
\rput(0, 3){\Rnode{a}{$\{P, Q\}$}}
\rput(4, 3){\Rnode{c}{$\{\neg P, Q\}$}}
\rput(6, 3){\Rnode{d}{$\{\neg P, \neg Q\}$}}
\rput(3, 2){\Rnode{f}{$\{Q\}$}}
\rput(4, 1){\Rnode{g}{$\{\neg P\}$}}
\ncline[nodesep=2pt]{a}{f}
\ncline[nodesep=2pt]{c}{f}
\ncline[nodesep=2pt]{d}{g}
\ncline[nodesep=2pt]{f}{g}
\endpspicture
\caption{Resolution DAG $D_1' = {\tt percolate}(D_1, A, L)$}
\label{fig8}
\end{figure}
\end{center}

Finally, we construct the resolution refutation
$D = {\tt graft}(D_1', D_2)$ shown in Figure \ref{fig9}:

\begin{center}
\begin{figure}[H]
\pspicture(0, 0)(7, 3)
\rput(0, 3){\Rnode{a}{$\{P, Q\}$}}
\rput(2, 3){\Rnode{b}{$\{P, \neg Q\}$}}
\rput(4, 3){\Rnode{c}{$\{\neg P, Q\}$}}
\rput(6, 3){\Rnode{d}{$\{\neg P, \neg Q\}$}}
\rput(1, 2){\Rnode{e}{$\{P\}$}}
\rput(3, 2){\Rnode{f}{$\{Q\}$}}
\rput(4, 1){\Rnode{g}{$\{\neg P\}$}}
\rput(2, 0){\Rnode{h}{$\square$}}
\ncline[nodesep=2pt,linewidth=1.5pt]{a}{e}
\ncline[nodesep=2pt,linewidth=1.5pt]{b}{e}
\ncline[nodesep=2pt]{a}{f}
\ncline[nodesep=2pt]{c}{f}
\ncline[nodesep=2pt,linewidth=1.5pt]{e}{h}
\ncline[nodesep=2pt]{d}{g}
\ncline[nodesep=2pt]{f}{g}
\ncline[nodesep=2pt,linewidth=1.5pt]{g}{h}
\endpspicture
\caption{Resolution DAG $D = {\tt graft}(D_1', D_2)$}
\label{fig9}
\end{figure}
\end{center}
where the edges coming from $D_2$ are indicated with thicker lines.

Observe that the proof of Theorem \ref{compP} proves that
if $\Gamma$ is unsatisfiable, then
our algorithm succeeds no matter which clause containing
at least two literals is chosen and no matter which literal
is picked in such a clause. 

Furthermore, as pointed out by one of
the referees, although the proof of completeness
is constructive in the
sense that it shows an algorithm is correct, it does not explicitly
use constructive logic. Nevertheless the logical proof can be recovered
from the algorithm and  it is constructive.

\section*{Acknowledgement}
The author wishes to thank Robert Constable and the referees
for very helpful comments.


\begin{thebibliography}{1}

\bibitem{ChangLee}
Chin-Liang Chang and Richard Char-Tung Lee.
\newblock {\em Symbolic Logic and Mechanical Theorem Proving}.
\newblock Academic Press, first edition, 1973.

\bibitem{Gallbooklogic}
Jean~H. Gallier.
\newblock {\em Logic For Computer Science}.
\newblock Wiley, first edition, 1986.

\bibitem{LewisPapa}
Harry~R. Lewis and Christos~H. Papadimitriou.
\newblock {\em Elements of the Theory of Computation}.
\newblock Prentice-Hall, first edition, 1981.

\bibitem{Robinson65}
J.A. Robinson.
\newblock A machine oriented logic based on the resolution principle.
\newblock {\em J.ACM}, 12(1):23--41, 1965.

\bibitem{Robinson79}
J.A. Robinson.
\newblock {\em Logic: Form and Function}.
\newblock North-Holland, first edition, 1979.

\bibitem{Underwood}
Judith Underwood.
\newblock The tableau algorithm for intuitionistic propositional calculus as a
  constructive completeness proof.
\newblock In Basin D., Fronhofer B., Hahnle R., Posegga J., and Schwind C.,
  editors, {\em Second Workshop on Theorem Proving with Analytic Tableaux and
  Related Methods, Marseille, France}, pages 245--248. Max--Planck--Institut
  fur Informatik, Saarbrucken, Germany, 1993.

\end{thebibliography}
\end{document}